\documentclass[preprint,12pt]{elsarticle}

\usepackage{amssymb}
\usepackage{graphicx}
\usepackage{subfigure}
\usepackage{wrapfig}
\usepackage{color} % To introduce colors by Editor 
  
\journal{Journal of Sound and Vibration}

\begin{document}

\begin{frontmatter}

\title{Effect of lateral confinement on the apparent mass of particle dampers}

\author[1,4]{Mar\'ia Victoria Ferreyra}\ead{ferreyravic@gmail.com}
\author[2]{Mauro Baldini}
\author[1,3]{Luis A. Pugnaloni}
\author[4]{St\'ephane Job}

\address[1]{Dpto. F\'isica, Facultad de Ciencias Exactas y Naturales, Universidad Nacional de La Pampa, Uruguay 151, 6300 Santa Rosa (La Pampa), Argentina.}
\address[2]{Departamento de Ingenier\'{\i}a Mec\'anica, Facultad Regional La Plata, Universidad Tecnol\'ogica Nacional, 60 esq. 124 S/N, 1900 La Plata, Argentina.}
\address[3]{Consejo Nacional de Investigaciones Cient\'ificas y T\'ecnicas, Argentina.}
\address[4]{Laboratoire Quartz EA-7393, Supm\'eca, 3 rue Fernand Hainaut, 93400 Saint-Ouen, France}

\begin{abstract}
We study, via Discrete Element Method simulations, the apparent mass ($m$, the ratio of a driving force to a resulting acceleration) and loss factor ($\eta$, the ratio of dissipated to stored energy) of particle dampers attached to a vertically driven, single degree of freedom mechanical system. Particle dampers (or granular dampers) consist in receptacles that contain macroscopic particles which dissipate energy, when they are subjected to vibration, thanks to the inelastic collisions and friction between them. Although many studies focus on $\eta$, less work has been devoted to $m$. The apparent mass of particle dampers is an important characteristic since the internal grains, which are free to move or collide inside their container, act as non-constant and time-dependent mass which alters the mass of the main vibrating system in a non-trivial way. In particular, it has been recently demonstrated [M. Masmoudi \textit{et al}. Granular Matter 18 (2016) 71.] that $m$ non-linearly depends on the driving acceleration $\gamma$ according to a power law, $m\propto\gamma^k$. Experiments using three-dimensional (3D) packings of particles suggest $k=-2$. However, simulations with one-dimensional (1D) columns of particles on a vibrating plate and theoretical predictions based on the inelastic bouncing ball model (IBBM) suggest that $k=-1$.
%, that fairly reproduce experiments ($k\simeq-2$, considering 3D random close packing of grains driven by a shaker) and models or simulations ($k\simeq-1$, considering a single inelastic bouncing ball model, IBBM, or a 1D vertical column of grains bouncing on kinematically driven plate). 
These findings left open questions whether the apparent mass, relying on how linear momentum is transferred from the damper to the primary system, may depend on the dimensionality of the packing or on lateral interactions between walls and grains. In turn, $\eta$ was shown to follow a universal curve, $\eta\propto\gamma^{-1}$, whatever the dimensionality and the constraints in the motion of the grains. In this work, we consider particle dampers without a lid under different confinement conditions in the motion of the particles (1D, quasi-1D, quasi-2D and full 3D). We find that the dynamical response of the granular damper ($m$ and $\eta$) is not sensitive to the lateral confinement or dimensionality. However, we have observed two distinct regimes, depending if the driving frequency is above or below the resonant frequency of the primary system. (i) In the inertial regime, $\eta$ decays according to the IBBM for all dimensions, $\eta\propto\gamma^{-1}$, while $m$ falls with an apparent power law behaviour that matches Masmoudi's experiments, $m\propto\gamma^{-2}$, for all dimensions but only in the range of moderate acceleration, before becoming negative for very high accelerations. (ii) In the quasi-static regime, both $m$ and $\eta$ display a complex behavior as functions of the excitation amplitude, but tend to the IBBM prediction, $m\propto\gamma^{-1}$ and $\eta\propto\gamma^{-1}$.
\end{abstract}


\end{frontmatter}

\section{Introduction}
\label{sec:introduction}

A particle damper (PD) is a device designed to increase the  damping of a vibrating system by inserting particles of macroscopic sizes in a receptacle attached to (or carved into the structure of) a primary system \cite{Panossian1992}. During vibration, the primary system transfers momentum to the particles which collide with each other and convert their kinetic energy into heat through the dissipative interactions (inelasticity and friction). This leads to a highly nonlinear response, but become an effective, inexpensive and durable damping mechanism in a wide range of frequencies \cite{Simonian1995,Xu2004}. A number of parameters such as size and shape of the particles, density, restitution coefficient, size and shape of the enclosure, and the type of excitation of the primary system, among others, are important in the PD performance \cite{Marhadi2005}.

Apart from dissipating energy and attenuating vibration, a PD alters the apparent mass of the primary system non-trivially. This is an important feature since the apparent mass has to be taken into account during design of the primary system. Yang \cite{Yang2003} has found that, under some conditions, the system may display effective (or apparent) masses above $M+m_\mathrm{p}$ or below $M$, where $M$ is the mass of the primary system and $m_\mathrm{p}$ is the total mass of the particles. More recently, S\'anchez et al. \cite{Sanchez2011} showed that, for a PD vibrated vertically, the effective mass can indeed over-shut and under-shut these two apparent limits (i.e., $M+m_\mathrm{p}$ and $M$) by fine tuning the gap left for the grains to move in the receptacle. Moreover, they found that for the optimal gap where the maximum attenuation is achieved, the effective mass matches the primary mass $M$. However, these previous studies defined the effective mass in terms of an overall parameter fitting the frequency response function (FRF) of the system over a broad frequency range. More recently, Masmoudi et al. \cite{Masmoudi2016} used a more natural definition for the apparent mass as the ratio of the Fourier components of the driving force and acceleration, for any specific vibration condition (i.e., excitation amplitude and frequency).

Masmoudi et al. \cite{Masmoudi2016} predicted the loss factor and apparent mass as functions of the vibration amplitude based on the Inelastic Bouncing Ball Model (IBBM) \cite{Araki1985,Metha1990,Pastor2007}. This prediction assumes that the mass of particles collides effectively as one singe ball with zero restitution on the enclosure floor. In the cases where these perfectly inelastic collisions occur periodically, the calculations can be simplified and solved analytically. Interestingly, both, the loss factor and the apparent mass, are predicted to decrease as a power law of the magnitude of the driving acceleration, with exponent $-1$. Although Masmoudi et al. \cite{Masmoudi2016} found that numerical simulations of one dimensional systems agreed with the predictions, experiments with a full three-dimensional enclosure provided conflicting results. The loss factor did show the $-1$ power law exponent, however, the apparent mass showed an apparent $-2$ power low decay. This opened the question as to whether the three-dimensional nature of a particle damper would make the prediction based on the IBBM unreliable for the apparent mass. Previous works suggested that 2D simulations provide correct predictions for the damping of real 3D dampers \cite{Sanchez2014}. More recently, Windows-Yule et al. \cite{Windows-Yule2017} showed that properties such as the stopping time for particles subjected to a tap in a container seem to be insensitive to the confinement of the particles. However, the effect of the confinement on the apparent mass has not been considered so far.

In this work, we focus our attention on the effect of the confinement on the apparent mass and on the loss factor of the particle damper. We use Discrete Element Simulations (DEM) which have been shown to be particularly useful in the past to study these systems \cite{Mao2004,Saeki2002,Bai2009,Fang2006,Sanchez2012}. In order to investigate the effect of the confinement, we consider different enclosures aspect ratios, which constrain the particles to move within a 1D, quasi-1D and quasi-2D spaces and compare results with a 3D case. In addition, the apparent mass relying on how the linear momentum and the energy are transferred from one system to the other, we also consider driving the primary system in both the inertial regime (i.e. with a driving frequency above the resonance frequency of the primary system) and in the quasi-static regime (low frequency limit, where the effect of the elastic spring of the primary system overcomes that of its mass).

In Sec. \ref{sec:model}, we present the system under study and in Sec. \ref{sec:dem}, we present the numerical scheme used to determine its mechanical response. In Sec. \ref{sec:analysis}, we remind the definition for the apparent mass and the loss factor. Finally, we investigate the regime of high and low frequency in Sec. \ref{sec:high_freq} and Sec. \ref{sec:low_freq}, respectively.

\section{The SDoF model}
\label{sec:model}

\begin{figure}[th]
  \begin{center}
     \includegraphics[width=0.5\textwidth]{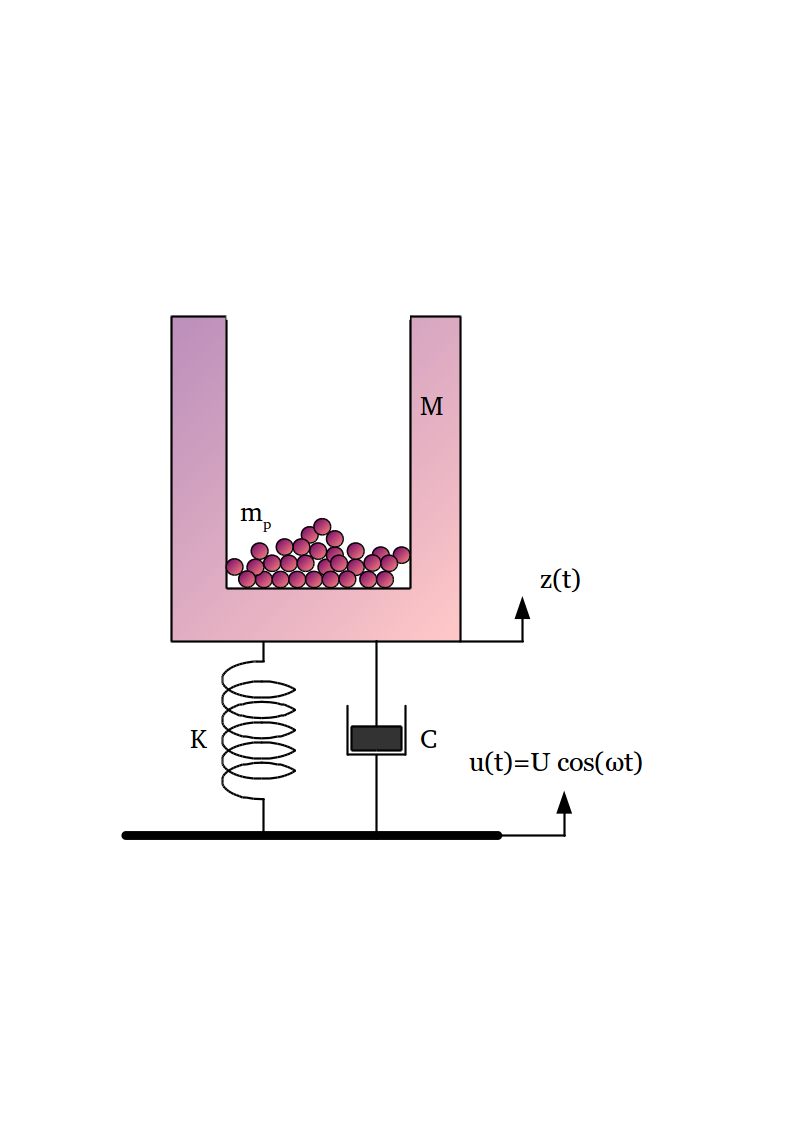}
  \end{center}
  \caption{Model of the SDoF system with a particle damper.}
  \label{fig:scheme}
\end{figure}

The single degree of freedom (SDoF) model with the PD is sketched in Fig. \ref{fig:scheme}. It consists in a primary system of mass $M=2.37$ kg, attached to a base by a spring of stiffness $K = 21500$ Nm$^{-1}$ and a small structural viscous damper with damping constant $C=7.6$ Nsm$^{-1}$. The primary system has the shape of an open box in which a number of spherical particles are deposited (see below for the details). The base is harmonically excited by moving its vertical position $u(t)$ as $u(t)=U \cos{\omega t}$, being $U$ the amplitude and $\omega$ the angular frequency of the base displacement. The gravitational acceleration is set to $g = 9.8$ ms$^{-2}$, and acts in the negative vertical direction. The natural frequency of the system for $C=0$ and without PD is $\omega_{0} = \sqrt{K/M}$ such that $f_{0}=\omega_{0}/2\pi\simeq15.16$ Hz. The equation of the motion of the primary system is
\begin{eqnarray}
M\gamma(t) &=& F(t) + F_\mathrm{p}(t),\label{eq:sdof-1}\\
F(t) &=& K[u(t)-z(t)]+C[\dot{u}(t)-\dot{z}(t)],\label{eq:sdof-2}\\
\gamma(t) &=& \ddot{z}(t),\label{eq:sdof-3}\\
u(t) &=& U\cos(\omega t),\label{eq:sdof-4}
\end{eqnarray}
where $z(t)$ is the position of the primary mass, $\dot{z}(t)$ and $\ddot{z}(t)$ are the first and second order time derivatives and $F_\mathrm{p}(t)$ is the vertical component of the total force exerted by the particles on the container inner walls. The system of Eqs. (\ref{eq:sdof-1}--\ref{eq:sdof-4}) thus stands for a single scalar differential equation of second order in $z(t)$, coupled to the dynamics of the particles at the colliding inner interfaces of the box. This set of equations can be solved numerically, see Sec. \ref{sec:dem}; the control parameters are $U$ and $\omega$. 

Although the movement of the primary system is restricted to the vertical direction, the particles are free to move in three dimensions. Initially, the particles are placed on the sites of an arbitrary BCC lattice: they quickly rearrange due to the gravity and the vibration. To simulate different confinement conditions, the lateral dimensions of the prismatic enclosure are set to: $1.1d \times 1.1d$ for the quasi-1D case,  $1.1d \times 16.1 d$ for the quasi-2D case, and $6.1d \times 6.1d$ for the 3D case, being $d$ the diameter of one particle. In this way, we confine particles to move within a narrow tube (1D) or a narrow slot (2D), or allow for a fully developed 3D motion (see Fig. \ref{fig:scheme}). The choice of lateral dimensions that are non-commensurated with the particle diameter prevents the formation of ordered patterns, especially at low vibrations intensities. In addition to the previous simulations, we carried out a set of 1D strict simulations as in \cite{Masmoudi2016}. These simulations are identical to the quasi-1D simulations with the exception that particles are aligned vertically, with normal interactions between particles only and no interact with lateral walls.

Here, it is important to understand that $F_\mathrm{p}(t)$ stands for an internal force, which is not accessible from a direct measurement: the PD has to be considered as a black box, with a non-constant mass, on which the primary structure applies the single external force $F(t)$ given in Eq. (\ref{eq:sdof-2}) to induce the acceleration $\gamma(t)$ given in Eq. (\ref{eq:sdof-3}). As a consequence, the PD results in a mechanical load that can be interpreted \cite{Masmoudi2016} as an apparent mass $m$ in addition to the mass of the container $M$, by reconsidering the Eq. (\ref{eq:sdof-1}):
\begin{equation}
[M+m(m_\mathrm{p},U,\omega,\dots)]\gamma(t) = F(t).\label{eq:sdof-app}
\end{equation}

Unraveling the functional dependency and the behavior of the apparent mass $m$, in particular in terms of magnitude and frequency of the driving vibration, is one goal of the present study.

\section{Discrete Elements Method}
\label{sec:dem}

We simulate the motion of spherical particles inside a prismatic box using a DEM approach \cite{Cundall1979,Poschel2005}. A collision between two particles $i$ and $j$, of radii $R_{i}$ and $R_{j}$ is modeled by normal and tangential components of the contact force that is applied at the contact point. The normal component is given by the Hertz--Kuwabara--Kono model \cite{Schafer1996,Kruggel12007}
\begin{equation}
    F_\mathrm{n} = -k_\mathrm{n}\alpha^{3/2}-\gamma_\mathrm{n}\upsilon_\mathrm{n}\sqrt{\alpha}, \label{eq:F_norm}
\end{equation}
where $\alpha = R_{i} + R_{j} - d_{ij}$ is the overlap between the spheres, being $d_{ij}$ the distance between the center of mass of the particles, $k_\mathrm{n}=(2E/3)\sqrt{R/2}(1-\nu^{2})^{-1}$ is the normal stiffness (with $E$ the Young's modulus, $\nu$ the Poisson's ratio and $R^{-1}=R_i^{-1}+R_j^{-1}$), $\gamma_\mathrm{n}$ the normal damping coefficient of the viscoelastic contact, and $\upsilon_\mathrm{n}$ the relative normal velocity.

The tangential component of the contact force takes the minimum value between a shear damping force and the dynamic friction \cite{Schafer1996,Kruggel22008}
\begin{equation}
 F_\mathrm{s} = -\min\left(\left|\gamma_\mathrm{s}\upsilon_\mathrm{s}\sqrt{\alpha}\right|,\left|\mu_\mathrm{d}F_\mathrm{n}\right|\right)\rm{sgn}\left(\upsilon_\mathrm{s}\right), \label{eq:F_shear}
\end{equation}
where $\gamma_\mathrm{s}$ is the shear damping coefficient, $\upsilon_\mathrm{s}$ the relative tangential velocity between the two spheres in contact and $\mu_\mathrm{d}$ the dynamic friction coefficient. We do not model static friction in these simulations. This means that conditions of mechanical equilibrium are poorly represented. However, since we are interested only in states of motion of the particles while the primary system is vibrated, this does not impose a limitation in our study.

Numerically, the equations of motion, including the degree of freedom of the container defined in Eqs. (\ref{eq:sdof-1}--\ref{eq:sdof-4}) and the degree of freedom of all the particles interacting via the forces given in Eqs. (\ref{eq:F_norm}--\ref{eq:F_shear}), are integrated using the velocity Verlet algorithm \cite{Allen1989}. Rotations are handled via quaternions \cite{Allen1989,Goldstein2002}.

The enclosure of the PD is simulated using five flat walls (a base and four lateral walls) with the same material properties as the particles. We focus our attention on the condition for which the particles never reach the ceiling of the container. Therefore, the enclosure can be considered infinitely high; consequently, its vertical dimension (height) becomes irrelevant in our study. 
\begin{table}[htb]
\centering
\begin{tabular}{cc}
\hline \hline Property & Value \\
\hline
Young's modulus $E$& $2.03\times10^{11}$ Nm$^{-2}$ \\
Density & 8030 kgm$^{-3}$ \\
Poisson's ratio $\nu$ & 0.28 \\
Friction coefficient $\mu_\mathrm{d}$ & 0.3 \\
Normal damping coefficient $\gamma_\mathrm{n}$ & $3.660\times10^{3}$ kgs$^{-1}$m$^{-1/2}$ \\
Shear damping coefficient $\gamma_\mathrm{s}$ & $1.098\times10^{4}$ kg$s^{-1}$m$^{-1/2}$ \\
Time step $\delta t$ & $8.75\times10^{-8}$ s \\
Time of simulation & 13.12 s \\
Particle radius $R$& 0.003 m \\
Total particle mass $m_\mathrm{p}$ & 0.227 kg \\ 
\hline \hline
\end{tabular}
\caption{Material properties of the particles and simulation parameters.}
\label{tab:parameters}
\end{table}

For the 3D and quasi-2D simulations we include $250$ particles in the enclosure with the material properties listed in Table \ref{tab:parameters}. Given the lateral span of the receptacles this corresponds to about $7$ (3D) and $15$ (2D) layers of particles. In the quasi-1D simulations we reduced the number of particles to $15$ in order to avoid a tall column as done by other \cite{Windows-Yule2015}. In this case, the low number of grains prevents a full dissipation of energy at each collision of the granular column  with the base (for a detailed discussion see \cite{Sanchez2012}). To prevent this effect the values of $\gamma_n$ and $\gamma_s$ were increased by a factor of ten. We also run simulations using a strict 1D system. In this case, particles move only along a vertical line and do not interact with the side walls. The rotational degree of freedom of the particles is also absent in the strict 1D case.

\section{Loss factor and apparent mass}
\label{sec:analysis}

We analyze the dynamical properties of the PD focusing on the loss factor $\eta$ and the apparent mass $m$ of the grains, as proposed by M. Masmoudi \textit{et al.} \cite{Masmoudi2016}. At a given driving frequency $\omega$, $\eta$ is calculated as the ratio between the energy dissipated per cycle $E_d$ and the maximum kinetic energy stored during a cycle $E_k$, 
\begin{equation}
\eta = E_d/2 \pi E_k = \tan \phi,\label{eq:loss}
\end{equation}
where $\phi =  \phi_z - \phi_F$ is the phase shift between the displacement $z(t)$ of the PD and the external force $F(t)$ exerted on the PD, see Eqs. (\ref{eq:sdof-2}) and (\ref{eq:sdof-3}). In practice, $\phi_z$ is conveniently estimated from the measured acceleration, $\phi_z = \phi_\gamma+\pi$. The phases $\phi_\gamma$ and $\phi_F$ and the magnitudes $A_F$ and $A_\gamma$ are the fundamental component of the Fourier series of the acceleration and force, respectively. These are obtained as
\begin{eqnarray}
A_F e^{j\phi_F} &=& \frac{1}{nT} \int_0^{nT} F(t) e^{-j\omega t} dt,\\
A_\gamma e^{j\phi_\gamma} &=& \frac{1}{nT} \int_0^{nT} \gamma(t) e^{-j\omega t} dt,
\end{eqnarray}
where $n$ is the number of cycles and $T=2\pi/\omega$ is the driving period. According to Eq. (\ref{eq:sdof-app}), the real-valued apparent mass of the particles, probed at the driving frequency, can be straightforwardly estimated from the ratio of the magnitudes of the fundamental components of the Fourier series of $F(t)$ and $\gamma(t)$, 
\begin{equation}
m = A_F / A_\gamma - M, \label{eq:meff}
\end{equation}
where $M$ is the mass of the container.  

\section{High frequency regime}
\label{sec:high_freq}

In this section we focus on the response of the system at a driving frequency ($f=160$~Hz) that is one order of magnitude above the natural frequency of the system ($f_0 \simeq 15.16$~Hz). This corresponds to the inertial regime, where the effect of the primary moving mass $M$ overcomes the elastic stiffness $K$, see Eqs. (\ref{eq:sdof-1}--\ref{eq:sdof-4}), $(M\ddot{z}\propto M\omega^2)\gg (Kz\propto M\omega_0^2)$. It is worth noticing that this configuration corresponds to the experimental case considered by Masmoudi et al. \cite{Masmoudi2016}, where the driving was above $166$ Hz and where the primary system is a medium sized shaker with a $10$ Hz typical resonant frequency.

\subsection{Results}

\begin{figure}[th]
\begin{center}
  \includegraphics[width=1\textwidth]{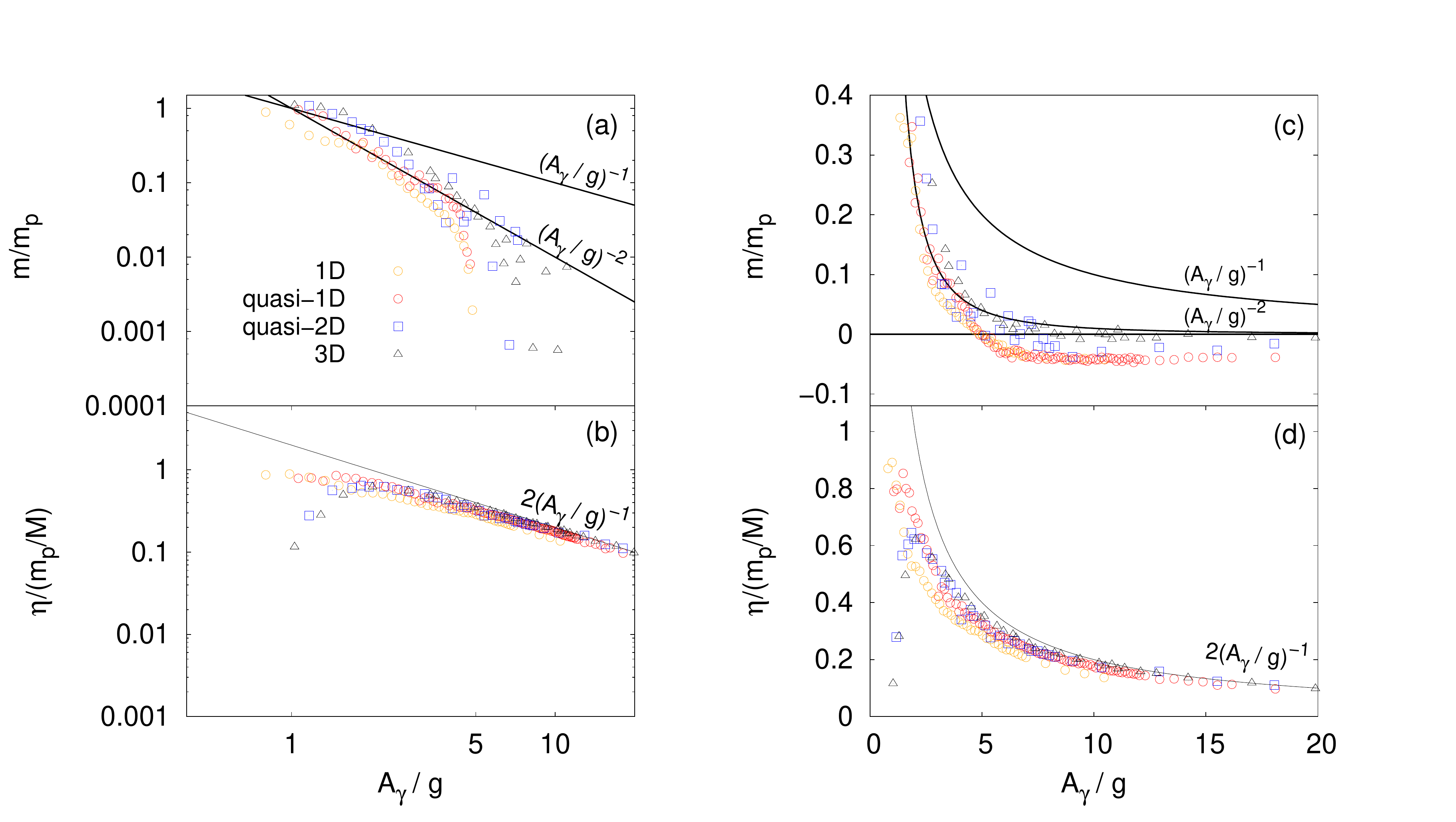}
\end{center}
\caption{(a) Apparent mass $m/m_\mathrm{p}$ of the grains for the high frequency regime ($160$ Hz) as a function  of $A_\gamma/g$ for 3D, quasi-2D, quasi-1D and 1D confinements. Due to the log--log scale, some data points for large $A_\gamma/g$ are missing because $m<0$. The black solid lines correspond to different possible integer power laws. (b) Loss factor $\eta$ scaled by the mass ratio $m_\mathrm{p}/M$ for the same simulations as in (a). The black solid line corresponds to the power law $2(A_\gamma/g)^{-1}$ predicted by Masmoudi \textit{et al.} \cite{Masmoudi2016}. (c-d) Linear plots for $m/m_\mathrm{p}$ and $\eta/(m_\mathrm{p}/M)$, respectively.}
\label{fig:160hz}
\end{figure}

Figure \ref{fig:160hz} shows the dependence of $m$ and $\eta$ as a function of the scaled driving acceleration $A_\gamma/g$ at an excitation frequency of $160$ Hz for 1D, quasi-1D, quasi-2D and 3D conditions. As we can see, simulations from different confinement conditions yield  similar values of apparent mass and loss factor. This seems to be in line with other studies of the effect of confinement where the stopping time for a granular column subjected to a tap is insensitive to the lateral confinement \cite{Windows-Yule2017}.  

If the amplitude of the acceleration $A_\gamma$ is below $g$, there is no relative movement between the particles. In this case, the system behaves as a rigid body: the granular material apparent mass tends to unity, $m/m_\mathrm{p}=1$, and there is no dissipation by the particles, $\eta=0$. We do not simulate the system for $A_\gamma < g$ because our simulation model does not take into account the static friction between the particles, which is essential for this static regime.

For $A_\gamma> g$, the grains detach from the base and collide with it intermittently. As $A_\gamma$ increases, the time during which the particles fly before colliding the base also increases. Since the lack of contact with the base makes the particles not to contribute to the mass of the system, $m$ falls as $A_\gamma$ is increased. The same is true for the loss factor since dissipation of energy occurs mainly during the collisions of the granular bed with the base. During these collisions the inelastic collapse caused by the large number of particle--particle collisions dissipate the entire kinetic energy of the grains \cite{Sanchez2012}.

When plotted in logarithmic scale, see Fig. \ref{fig:160hz}(a), $m$ appears to follow a power law $m \propto (A_\gamma/g)^{k}$ with $k\simeq -2$ for all confinement conditions. Strikingly, this observation is consistent with the 3D experimental finding obtained by Masmoudi \textit{et al.} \cite{Masmoudi2016}. However, in Ref \cite{Masmoudi2016}, 1D strict simulations and an analysis based on IBBM both gave a power law $m \propto (A_\gamma/g)^{-1}$. In this previous study, it is worth mentioning that the container is not considered as a degree of freedom in both simulation and model: its displacement, velocity and acceleration are enforced. In turn, the experiments were performed using a shaker, whose resonant frequency is generally considered as low enough to be negligible for the frequency range of interest. Similarly, our 1D simulations accounts for the container as a degree of freedom via its support (spring and vibrating base), and the results match the experimental observations. In light of these findings, one understands that the power law exponent $k\simeq-2$ observed in Ref. \cite{Masmoudi2016} thus relies on the response of the primary system to the impacts of the particles, and in particular on how linear momentum is transferred from one to the other during the collisions.

When plotted in linear scale, while extending the driving to larger acceleration, see Fig. \ref{fig:160hz}(c), one observes that the apparent mass becomes negative, $m<0$ above $A_\gamma/g \gtrsim 6$. Strikingly, this means that the apparent mass of the whole system (container and particles) is below the original mass of the container alone. As a consequence, describing the behavior of the apparent mass in terms of a power law appears to hold for moderate driving acceleration only, $A_\gamma/g < 6$. Above this threshold, a subtle difference can be appreciated for the different confinements: as one moves from 1D to 2D and 3D cases, the negative values of $m$ decrease in absolute value.

The loss factor $\eta$ appears to follow a well defined universal curve for all confinement conditions, in close agreement with the power law  $\eta = 2 (m_\mathrm{p}/M) (A_\gamma/g)^{-1}$ predicted in \cite{Masmoudi2016} for $A_\gamma/g \gtrsim 3$. Here, our results agree with this prediction (see Figs. \ref{fig:160hz}(b,d)) whatever the confinement conditions, and are also consistent with the experimental observations given in \cite{Masmoudi2016}.

\subsection{Discussion on negative apparent masses}

\begin{figure}[t!] 
 \begin{center}
  \includegraphics[width=0.95\textwidth]{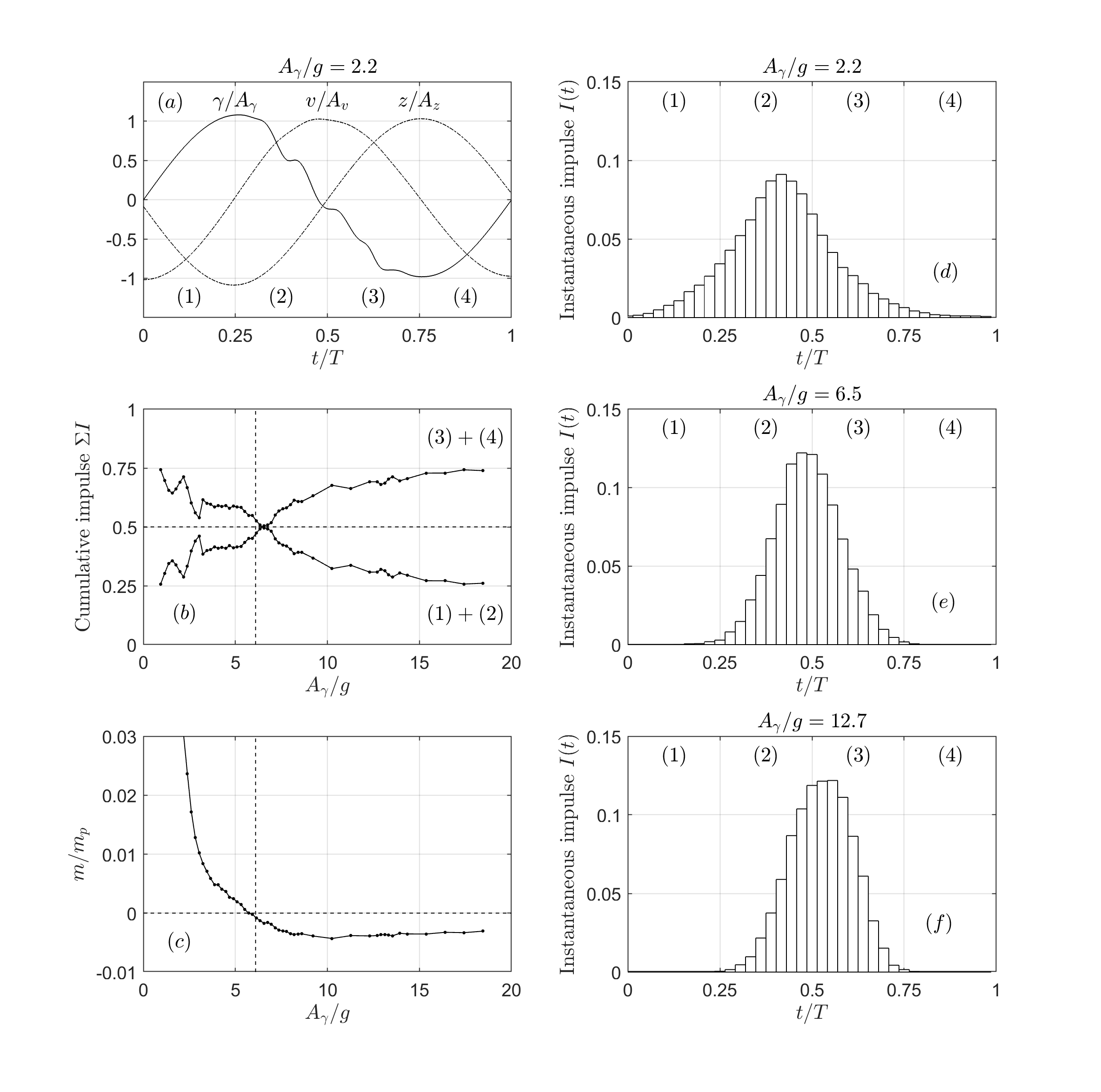}
\end{center}
\caption{(a) Sample position $z$, velocity $v$ and acceleration $\gamma$ of the box scaled by the corresponding mean amplitude of the signal as a function of time along one excitation period. The period has been divided in four quarters indicated as (1), (2), (3) and (4). (b) Fraction of the impulse transferred by the particles to the container in parts (1)+(2) and parts (3)+(4) of the cycle (the results accounts for 2000 cycles). (c) Apparent mass $m/m_{\mathrm{p}}$ of the grains as a function of the acceleration for a 1D system. The dashed lines indicate the point where the crossover between the contributions of part (1+2) and (3+4) occurs. (right column) Instantaneous impulse $I(t)=\int_{t}^{t+\delta t}{F_p(t)dt}$ as function of time within a cycle, gathered over 2000 periods, being $\tau\ll T$ an arbitrary small time span. Examples shown are (d) below the crossover $A_\gamma/g=2.2$, (e) at the crossover $A_\gamma/g=6.5$ and (f) above the crossover $A_\gamma/g=12.7$.}
\label{fig:negativem}
\end{figure}

As we have pointed out, for $A_\gamma/g \gtrsim 6$, the apparent mass $m$ of the grains becomes negative (i.e., the apparent mass of the total system falls below the mass $M$ of the primary system). This has been previously described for PD where collisions with base and ceiling are allowed \cite{Sanchez2011}. In our case, this negative $m$ is observed when, in average, the grains collide with the base in the portion of the cycle when the force of the spring is negative (i.e., when the position of the container is above the equilibrium position). Under this condition, the collision of the granular sample with the floor occurs when the box is being pulled downwards by the external force of the spring, which ``helps''  the external force and this translates into a lower apparent mass of the system.

To quantify the mechanism described above for the occurrence of negative apparent masses, we extract the phase at which the collisions occur and the impulse that is imparted by the grains to the container. In Fig. \ref{fig:negativem}(a) we show an example of a single period of the signals of acceleration, velocity and position of the container from one of our 1D simulations. The collision of the particles are seen as perturbations on the acceleration signal, in a time region labeled $(2)$ and $(3)$ where the velocity of the container is positive. Taking into account that the velocity of the particles are always downward just before a collision, this time region corresponds to head-on frontal impact. It is worth mentioning that impacts in the time regions $(1)$ and $(4)$ certainly occur, but in this case the smaller relative velocity (both the container and the particles go downward) yields a "soft landing" of the particles (with same velocity as the container) that generates weaker impulses that do not participate efficiently in the transfer of momentum from one to the other.

Quantifying how and when linear momentum is transferred from the particles to the container can be achieved by first estimating the instantaneous impulse $I(t)=\int_{t}^{t+\tau}{F_p(t)dt}$, where $\tau$ is an arbitrary small time span $\tau\ll T$. The dynamics of the vibrated particles appears to be non-stationary over just few periods, so one can obtain a reliable statistics by gathering the impulse over 2000 cycles, i.e. by representing it as a function of time modulo the period. Three examples are shown in Fig. \ref{fig:negativem}(d-f), at small, intermediate and large driving accelerations. On purpose, these three examples are chosen such that they correspond to positive ($m>0$), null ($m\simeq0$), and negative ($m<0$) apparent mass, respectively. The comparison of the small and the large acceleration clearly shows that when the collision occurs more probably within the first time region $(1-2)$ then the apparent mass appears positive, while in the last time region $(3-4)$ the apparent mass appears negative. At the crossover, $\gamma/g\simeq6$, one finds approximately the same amount of impulse transferred in one or the other region, see Fig. \ref{fig:negativem}(c). The crossover can be unraveled in details by plotting the impulse accumulated in each of the four time regions as a function the driving acceleration, see Fig. \ref{fig:negativem}(b).

Considering that only the collisions within time regions $(2-3)$ of positive container's velocity participate to an efficient transfer of momentum, we thus find that when the container is downward (region $(2)$, i.e. when the force $F(t)$ pushes it up), then the collisions decelerate the container making it harder to move up. This situation is interpreted as a heavier degree of freedom than the original container, i.e. a positive additional apparent mass $m>0$. In contrast, collisions occurring in the time region $(3)$ decelerate the damper in a phase where the container is up and is already being decelerated due to the negative force $F(t)<0$ that pulls it down. Therefore, the impulse transferred in the downward direction aids the spring force in driving the system down, resulting in a negative apparent mass ($m<0$). As a rule of thumb, the asymmetric probability of collisions makes the apparent mass being negative or positive. Such an asymmetry is a consequence of the driving mechanism, which here comes from the action of a force on a resonant system.

The effect of the collision phase could not be seen in Masmoudi's simulations \cite{Masmoudi2016} since the container displacement, velocity and acceleration were enforced kinematically. This situation results in neglecting the dynamical response of the container to impacts, and thus part of the coupling between the particles and the container. This coupling likely introduces a phase shift that produces the aforementioned positive or negative apparent mass. In order to probe these assertions, we consider in the following section the response of our damper in the quasi-static regime, which likely corresponds more closely to the configuration considered in \cite{Masmoudi2016} for the 1D simulations and the IBBM.

\section{Low frequency regime}
\label{sec:low_freq}

In this section we focus on the response of the system in the quasi-static regime, where the inertia of the container can be neglected in comparison to the elastic restoring force of the primary system spring, $(M\ddot{z}\propto M\omega^2)\ll (Kz\propto M\omega_0^2)$. It is worth noticing that this configuration likely correspond to the configuration simulated by Masmoudi et al. \cite{Masmoudi2016}, where the kinematical driving corresponds to an infinitely stiff primary system which enforce a displacement. Here, the driving frequency is about half of the natural frequency of the primary system, $f=8\mbox{ Hz } \simeq f_0/2$, which fulfills the condition of the quasi-static regime at lowest order.

\subsection{Results}

\begin{figure}[th]
\centering
 \includegraphics[width=0.7\textwidth]{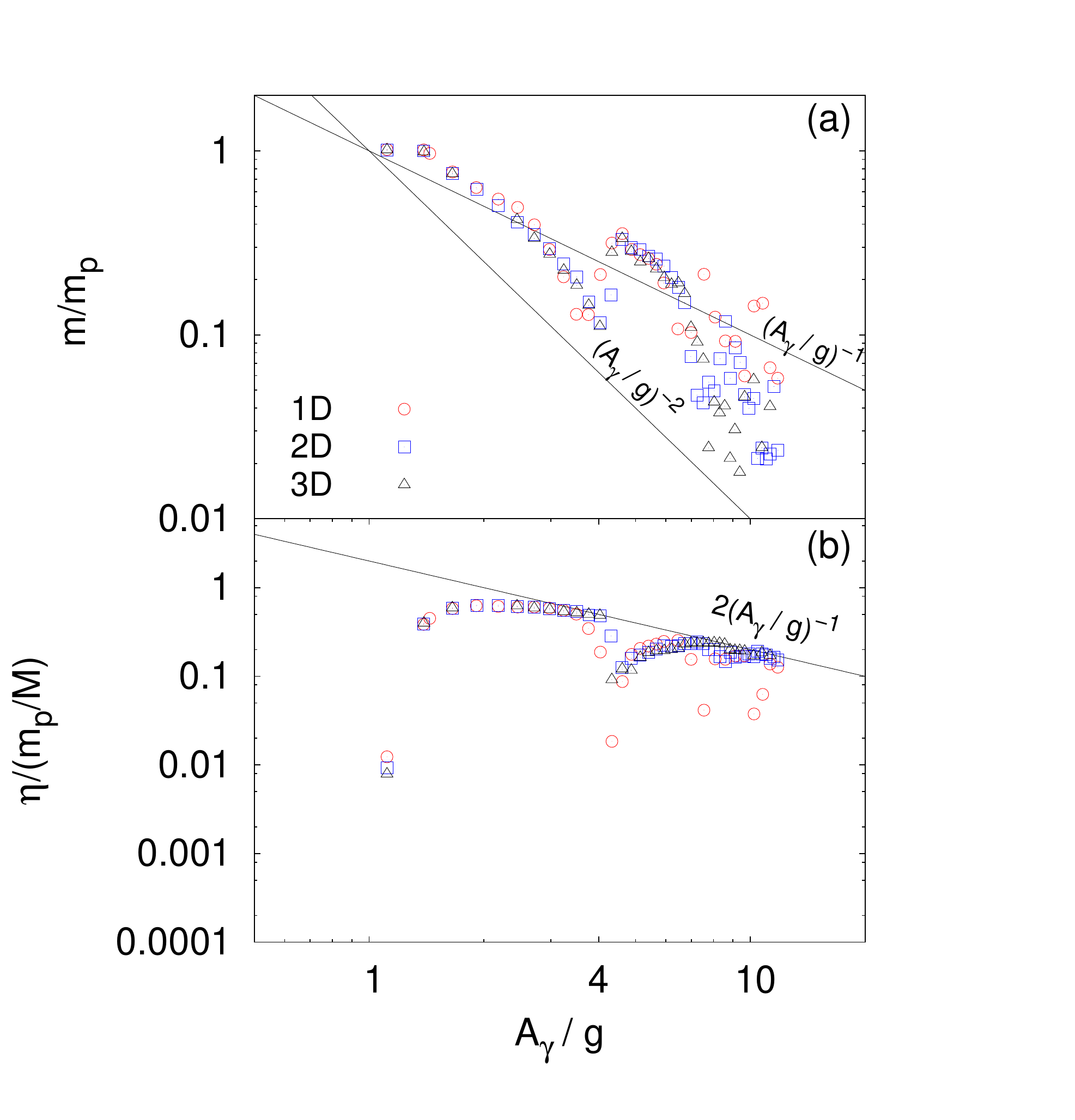}
 \caption{(a) Apparent mass $m/m_\mathrm{p}$ of the grains for the low frequency regime ($8.0$ Hz) as a function of $A_\gamma/g$ for quasi-1D, quasi-2D and 3D simulations. The black solid lines correspond to different possible integer power laws. (b) Loss factor $\eta$ as a function of $A_\gamma/g$ for the same simulations. The black solid line corresponds to $\eta \propto 2 (M/m_\mathrm{p})(A_\gamma/g)^{-1}$. For sake of comparison, the straight lines $(A_\gamma/g)^{-1}$ shown in (a) and (b) stand for the IBBM prediction given in \cite{Masmoudi2016}, while the straight lines $(A_\gamma/g)^{-2}$ is an arbitrary ansatz.}
\label{fig:8hz}
\end{figure}

In Fig. \ref{fig:8hz}, we show the apparent mass and loss factor as a function of the acceleration for a low excitation frequency ($8.0$ Hz) using quasi-1D, quasi-2D and 3D systems. As we can see, for this low frequency value, all conditions of confinement yield the same values of $m/m_\mathrm{p}$ and $\eta$. This confirms the fact that lateral confinement does not affect the dynamics. However, $m/m_\mathrm{p}$ and $\eta$ display a much richer behavior in comparison with the high frequency case described in the previous section. Both, $m/m_\mathrm{p}$ and $\eta$, present sharp transitions at particular values of $A_\gamma/g$. Without loss of generality, let us focus on $A_\gamma/g \simeq 4.0$. For $A_\gamma/g \lesssim 4.0$, the apparent mass decreases with $A_\gamma$ roughly as $(A_\gamma/g)^{-2}$, while $\eta$ falls as $(A_\gamma/g)^{-1}$. At $A_\gamma/g \simeq 4.0$, $m/m_\mathrm{p}$ suddenly grows and $\eta$ drops. Right after these significant jumps both variables tend to retake the original trend: $m$ decreases and $\eta$ increases. Eventually, this behavior is repeated at $A_\gamma/g \simeq 6.0$, and $10.0$. These features are not observable at high frequencies. Interestingly, we do not observe a negative apparent mass $m<0$ in this low frequency limit.

\begin{figure}
 \includegraphics[width=1.0\columnwidth]{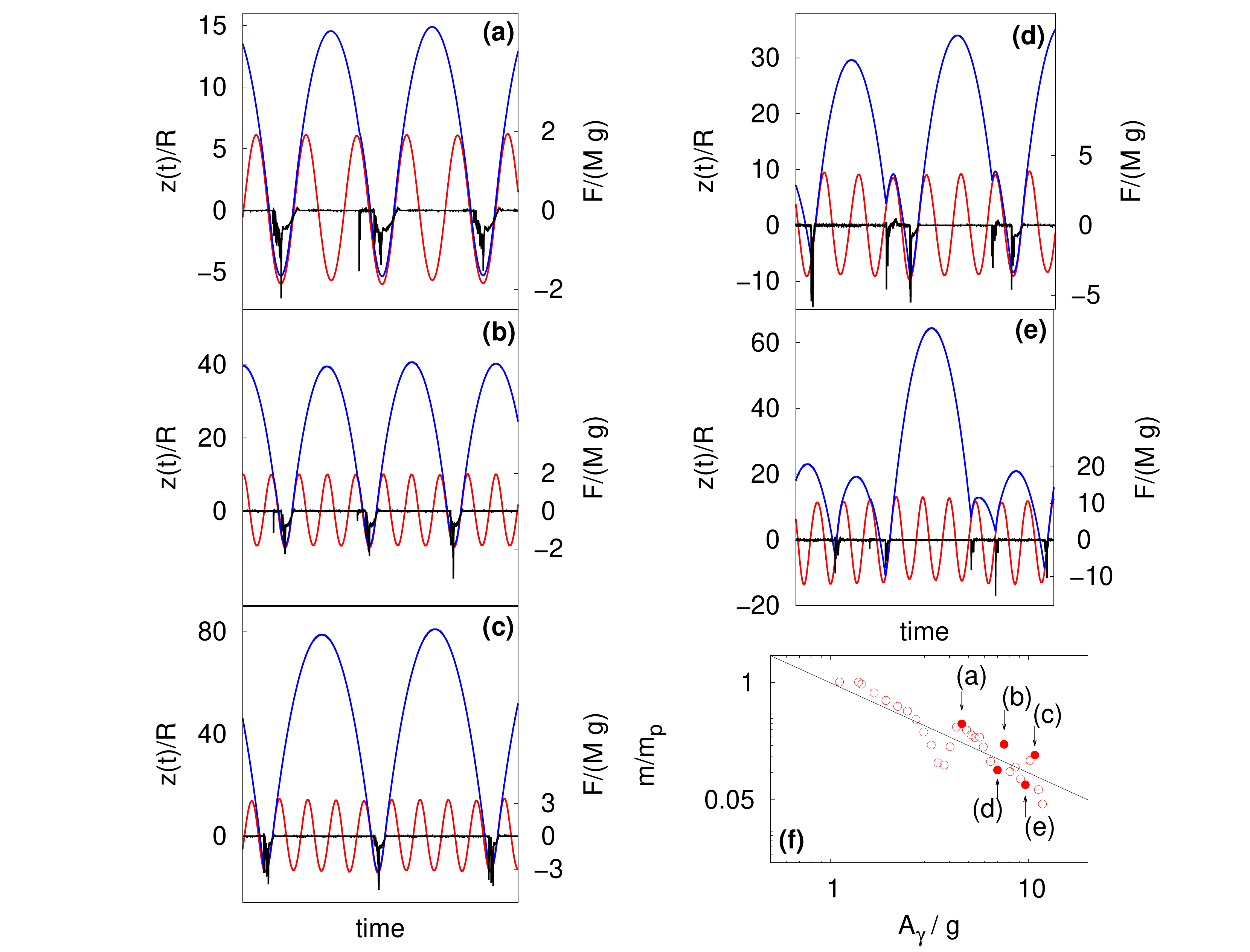}
 \caption{Trajectory of the enclosure floor (red line) and the lowest particle (blue line) in the granular sample as a function of time for various amplitudes $A_\gamma/g$: (a) $4.61$, (b) $7.56$, (c) $10.78$, (d) $6.99$ and (e) $9.66$. The simulations correspond to the quasi-1D system exited at $8.0$ Hz. The black line corresponds to the force (right axis) exerted by the particles on the floor. The time scale is different in each plot in order to display different number of cycles. In (a), (b) and (c) the particles collide every an entire number of periods and the impact occurs at zero relative velocity. (f) Apparent mass as in Fig. \ref{fig:8hz} for the quasi-1D system only. The arrows indicate the panel in which a sample trajectory of the simulation is shown.} \label{fig:tray_8hz}
\end{figure}

It is interesting to note that the upper bound for $m/m_\mathrm{p}$ and $\eta$ correspond to $(A_\gamma/g)^{-1}$, which coincides with the IBBM prediction made by Masmoudi et al. \cite{Masmoudi2016}. In Fig. \ref{fig:tray_8hz}, we show the trajectories of the enclosure floor (red line) and the lowest particle (blue line) in the granular sample as a function of time for various amplitudes $A_\gamma/g$: (a) $4.61$, (b) $7.56$, (c) $10.78$, (d) $6.99$ and (e) $9.66$. The black line represents the force (right axis) exerted by the particles on the floor, showing the exact moment of each collision. The chosen values of $A_\gamma/g$ correspond to the solid circles on panel (f). Panels (a), (b) and (c) correspond to points that fall on the $(A_\gamma/g)^{-1}$ power law. For these, we observe that the particles collide periodically with the base. This is incidentally an assumption made in Ref. \cite{Masmoudi2016} (see next section for a detailed discussion). Panels (d) and (e) correspond to the local minima in the $m$ curve. For these points we observe non-periodic trajectories.

\subsection{Discussion on apparent mass for periodic  and non-periodic responses}

In Ref. \cite{Masmoudi2016}, the authors estimated the loss factor and apparent mass of the real system using the IBBM \cite{Araki1985,Metha1990,Pastor2007}. For this, it is assumed that the particles collide periodically with the enclosure floor. Moreover, each collision are considered effectively as instantaneous and having a zero restitution coefficient, according to the IBBM. Within these approximations, one finds the loss factor
\begin{equation}
    \eta = (m_\mathrm{p}/M)\times(g/2A_\gamma)\times[1+\cos{(\arcsin{(g/A_\gamma)}+(2A_\gamma/g))}]^2.
    \label{eq:lossfactor_ibbm}
\end{equation}

The expression given in Eq. (\ref{eq:lossfactor_ibbm}) is valid for $A_\gamma/g\gtrsim3$ and tends to zero for few specific values, $A_\gamma/g\simeq 4.6$, $7.8$, $10.6$... where the particle lands on the container with zero relative velocity. The later case finely relies on precise collision phases, where no dissipation occurs, and these dips tend to disappear as long as one consider a finite duration for the contact dynamics of more than one particle on the container \cite{Job2007}. If so, the loss factor can be reliably approximated by its upper bound value,
\begin{equation}
    \eta = (m_\mathrm{p}/M)\times(2g/A_\gamma). \label{eq:lossfactor_approx}
\end{equation}

Concerning the apparent mass, the Masmoudi's prediction \cite{Masmoudi2016} is based on the assumption that the perfectly inelastic impacts occurs periodically, with a constant periodical time of flight. In this way, the IBBM predicts that
\begin{equation}
    (m/m_\mathrm{p}) = 1 \mbox{ for } A_\gamma<g \mbox{ and } (m/m_\mathrm{p}) \propto (A_\gamma/g)^{-1}\mbox{ otherwise}
    \label{eq:meff_ibbm}
\end{equation}

Here, Fig. \ref{fig:tray_8hz} shows that the IBBM prediction fails if the dynamics of the collisions becomes non-periodic, see for instance Fig. \ref{fig:tray_8hz}(d) and (e). This is a consequence of the proximity of a bifurcation point \cite{Metha1990,Pastor2007}, where the system experiences period doublings. In this case, the transfer of momentum is not synchronized with the driving, spreads over many time scales, and consequently becomes less efficient at the driving frequency. In this way, as for the loss factor, the IBBM prediction given in Eq. (\ref{eq:meff_ibbm}) stands only for an estimate of the upper bound.

\section{Conclusions}
\label{sec:conclusion}

We have studied the loss factor and apparent mass of a vertically vibrated particle damper as a function of the excitation energy for a low and a high frequency in comparison with the natural frequency of the primary system. We have considered particle dampers where particles are laterally confined in a quasi-1D or in a quasi-2D container as well as in a full 3D box. We have found that the lateral confinement does not affect neither the loss factor nor the apparent mass of the PD. 

We demonstrated that at high frequencies the loss factor complies with the theoretical prediction made by Masmoudi et al. \cite{Masmoudi2016} that $\eta =2 (m/M) (A_\gamma/g)^{-1}$ for all confinement conditions. However, the apparent mass does not seem to follow a clear power law as suggested by the experiments \cite{Masmoudi2016}. The apparent power law that we observe in a narrow range of accelerations is lost at high acceleration due to the negative values obtained for the apparent mass. 

At low frequencies, a regime that was not explored experimentally in \cite{Masmoudi2016}, we find that the apparent mass and the loss factor present very complex responses as a function of $A_\gamma/g$. However, the values of $\eta$ and $m$ have an upper limit compatible with the power laws predicted by Masmoudi et al. \cite{Masmoudi2016}. 

Although previous studies demonstrated that the damping efficiency of a particle damper is not affected by its dimensionality \cite{Sanchez2014,Windows-Yule2017}, there were no clear evidences that this also applies to the apparent mass of the system. Our results confirm that to predict both the apparent mass and the loss factor of a PD, 1D and 2D models are in fact suitable. 

The loss factor as a function of the acceleration is well predicted at high frequencies by the simple power law proposed by Masmoudi et al. \cite{Masmoudi2016}, and can then be used to design a PD. However, at low frequencies the power law seems to provide only an upper limit to both the apparent mass and the lost factor.

\section*{Acknowledgments}
We are grateful to Mart\'in S\'anchez for help with the simulations. This work was funded by ANPCyT (Argentina) through grant PICT-2016-2658. The first author (MVF) is grateful for the financial support of both Supm{\'e}ca and Campus France during her stay in France, at the laboratoire Quartz EA-7393.

\bibliographystyle{elsarticle-num}

\end{document}